\documentstyle[12pt,aasms4,psfig]{article}
\newcommand{\etal}{{\it et al.} }
\newcommand{\asca}{{\it ASCA} }

\newcommand{\exosat}{{\it EXOSAT} }
\newcommand{\ginga}{{\it Ginga} }

\newcommand{\src}{3C~273 }
\newcommand{\bsax}{{\it BeppoSAX} }
\newcommand{\rxte}{{\it RXTE} }
\newcommand{\fekalfa}{$\rm Fe~K\alpha$ }

\begin{document}

\title{A BROAD \fekalfa EMISSION LINE IN THE X-RAY SPECTRUM 
OF THE QUASAR \src}

\author{Tahir Yaqoob\altaffilmark{1,2},
Peter Serlemitsos\altaffilmark{1}}

\vspace{3cm}

\altaffiltext{1}{Laboratory for High Energy Astrophysics, 
NASA/Goddard Space Flight Center, Greenbelt, MD 20771, USA.}
\altaffiltext{2}{Joint Center for Astrophysics, Physics Department,
University of Maryland,
Baltimore County, Baltimore,
1000 Hilltop Circle, MD21250.}

\begin{center}
{\it Accepted for Publication in the Astrophysical Journal Letters 
(9/25/00; submitted 8/9/2000)}
\end{center}

\begin{abstract}
The \fekalfa line, an important physical diagnostic in the
X-ray spectra of active galaxies (AGNs) and quasars, has been 
notoriously difficult to measure  
in the high-luminosity, radio-loud quasar \src ($z= 0.158$).
On the few occasions that it has been detected its intrinsic width
has been thought to be narrow (FWHM $<10,000 \rm \ km \ s^{-1}$)
with an equivalent width (EW) of a few tens of eV. This was consistent
with the general trend that as one goes from low to high luminosity
AGNs the \fekalfa line goes from being strong 
(${\rm EW} \sim 200-300$ eV) and broad (FWHM $\sim 10^{5}$ km/s)
to being weak and narrow, or absent altogether. 
Here we present the results of new \asca and \rxte observations,
together with archival \asca data, and show
for the first time
that the \fekalfa line in \src is as broad as that seen in Seyfert 1
galaxies. The line is resolved in two of the observations, with a mean
Gaussian width of $0.8 \pm 0.3$ keV, corresponding to a FWHM of
$\sim 0.3 \pm 0.1c$. 
The smallest and largest equivalent width measured is $43_{-34}^{+34}$
eV and $133^{+52}_{-53}$ eV respectively (quasar frame).
The  Compton-reflection continuum is less
than 10\% of that expected from a centrally illuminated semi-infinite,
face-on, Compton-thick disk, confirming previous studies that 
Compton reflection is negligible in 3C 273. 
The largest values of the \fekalfa line EW are under-predicted if the
line originates in the disk, unless a time lag longer than 
several days between
line and continuum and/or an over-abundance of Fe
is invoked.
We cannot 
unambiguously constrain the disk inclination angle. About $60^{\circ}$ 
is preferred for a cold disk, while a face-on disk is allowed if
the ionization state of Fe is H-like.

\end{abstract}

\keywords{galaxies: active -- quasars: emission lines -- 
quasars: individual: \src
-- X-rays: galaxies}

\section{INTRODUCTION}

The {\it Advanced Satellite for Cosmology and Astrophysics}
(\asca -- Tanaka, Inoue, and Holt 1994) found that the  \fekalfa
fluorescent emission line in Seyfert~1 galaxies is
often very broad
and the data are 
consistent with an interpretation of 
an origin from 
cold matter
in an
accretion disk rotating around a central black hole
(see Fabian \etal 2000 and references therein).
The line profile is sculpted by characteristic
gravitational and Doppler energy shifts.
Currently, study of the  \fekalfa emission line is
the only way to probe
matter within a few to tens of gravitational radii
of a black hole.
These emission 
features 
become
less common 
in active galactic nuclei (AGNs) 
with a 2--10 keV intrinsic luminosity exceeding $\sim 10^{45}
 \rm \ erg \ s^{-1}$ (or synonymously, quasars). 
On the few occasions that an \fekalfa line is detected in high-luminosity AGN
it appears to be narrow (FWHM $< 10,000$ km/s) with a peak 
energy indicative of highly-ionized matter.  
This trend
is discussed in Nandra \etal (1997a,b), 
Reeves \etal (2000) and George \etal (2000)
and references therein. 

The bright, high-luminosity 
($L(2-10 \rm \ keV) \sim 0.8-2 \times 10^{46} \rm \ erg \ s^{-1}$;
$H_{0} = 50 \rm \ km \ s^{-1} \ Mpc^{-1}$ and $q_{0}=0$ throughout)
radio-loud quasar
\src
($z=0.158$) has been extensively studied in all wavebands
(e.g. see von Montigny \etal 1997; T\"{u}rler \etal 1999 and references therein).
In particular, X-ray studies have found
a featureless nonthermal continuum, with the occasional detection
of an \fekalfa emission line. Proportional counter experiments such
as those on \exosat and \ginga could not constrain the width of
the line and, for a narrow line assumption, found equivalent widths of 
$\sim 10-50$ eV when it was detected (e.g. Turner \etal 1990;
Williams \etal 1992). Subsequent observations with \asca 
again
found some weak detections with an equivalent width of the order
of tens of eV (Yaqoob \etal 1994; Cappi \etal 1998; Reeves \etal 2000).  
Observations with \bsax told a similar story
(Grandi \etal 1997; Orr \etal 1998; Haardt \etal 1998).

In these \asca and \bsax studies, the \fekalfa line in 3C 273 has been 
assumed or claimed to
be narrow. 
However, on two  
occasions (1996 July and 1998 June), 
during  coordinated {\it ASCA}, {\it RXTE} and {\it BeppoSAX} 
multi-mission cross-calibration campaigns,
we found that the \fekalfa line was in fact broad.
This is not 
inconsistent with previously published results because 
when the equivalent width of the line is small it is
very difficult to constrain the intrinsic width.
Results from the 1996 July and subsequent \asca observations
have not yet been published. 
As for the \bsax results (Grandi \etal 1997; Haardt \etal 1998),
recalibration of the
{\it MECS} aboard \bsax has recently brought 
the narrow width 
of the line into question 
(Grandi, private communication).
In this {\it Letter} we re-examine the old \asca data,
and present results from three new \asca observations, plus
two simultaneous RXTE observations, in order to investigate the
nature of the elusive \fekalfa emission line in this source and we
conclude that it is broad, just like that found in Seyfert 1 galaxies.
 
\section{THE DATA AND SPECTRAL FITTING RESULTS}

Between 1993, June and 2000, January, there were a total of
six \asca campaigns on 3C 273, resulting in twelve pointed
observations. Two observations have simultaneous \rxte data
(1996 July and 1998 June).
Table 1 shows an observation log, exposure times and count rates 
for these observations.
We utilize all four instruments aboard \asca
(SIS and GIS), fitting the same model to all detectors simultaneously
with only the relative normalizations independent.
All the SIS data were taken in 1-CCD mode. The \asca data
were analysed using the methods described in detail in Yaqoob \etal 1998.
Note that observation B was a calibration observation performed in
four parts, placing the source sequentially on each of the four SIS 
chips over a period of $\sim 5$ days. 
Observation C was a multiple `snapshot' science campaign
carried out over $\sim 11$ days. 
Preliminary spectral fitting showed that we could not obtain
useful constraints on the \fekalfa line parameters from the separate 
pointings due to insufficient signal-to-noise so we combined the spectra
from individual pointings for each of the data sets ``B'' and ``C''.
For \rxte we used only the 
Proportional Counter Array (PCA) data, which were reduced using methods
described in Weaver, Krolik, \& Pier (1998), except that we used a 
later version
of the spectral response matrix (V 7.01) and
background model (`L7\_240\_FAINT'). Only data from PCU 0,1 and 2,
layer 1 were used in this analysis.
Note that cross-calibration studies show that in the 2--10 keV band
PCA fluxes are systematically at least 20\% higher than \asca whereas
\asca and \bsax fluxes agree to a few percent
\footnote{ASCA GOF calibration memo at \\
http://lheawww.gsfc.nasa.gov/~yaqoob/3c273/3c273\_results.html}. We
will 
rely on the \asca fluxes for consistency since we discuss only two
\rxte observations.

Since we are only interested in the \fekalfa line properties for this work,
we only utilize data in the 1.5--10 keV band for \asca  
and 3--10 keV
band for the PCA (the latter does not have a useful response below 3 keV).
First, we fitted each data set with a simple power-law plus Galactic
absorption ($N_{H} = 1.79 \times 10^{20} \rm \ cm^{-2}$; 
Dickey \& Lockman 1990),
but omitting data between 4--8 keV (the Fe-K line region). Visual inspection
of the data/model residuals over the whole energy range
then revealed, in most cases, a hump peaking around 5.5 keV (observed frame),
presumably due to an \fekalfa emission line. The most dramatic of these was
for observation D, for which the SIS, GIS and
PCA data/model ratios are shown in Figure 1. 
It can be seen that the \fekalfa line is very broad, extending from
$\sim 4-7$ keV similar to the broad \fekalfa lines in Seyfert 1 galaxies. 
We emphasize that this is not a result of instrument calibration
problems. The line is broad as measured by five
independent instruments, out of which three instrument
groups (SIS, GIS and PCA) were calibrated independently in the 
relevant energy band.
We show below that even the PCA, despite
its poor energy resolution resolves the \fekalfa line at 
90\% confidence.
Before discussing quantitative measurements of the
line parameters we show in Figure 2
similar data/model ratios resulting from co-adding all the SIS0 and SIS1
data in Table 1. This constitutes {\it all} the SIS data 
for 3C 273 taken over the entire lifetime of {\it ASCA},
amounting to an exposure time of $\sim 258$ ks. The broad  \fekalfa line
can be seen with very high signal-to-noise. 
It is not meaningful
to perform more detailed spectral fitting on the co-added spectra 
because it is known that the spectral index of the continuum is variable
over long timescales (Turner \etal 1990)
so we will not utilize the co-added spectrum further. Note that in Figure 2,
at the highest energies (above $\sim 8$ keV), the SIS1 data are
systematically higher than SIS0. This is a known calibration uncertainty
which shows up in other sources when the signal-to-noise is high
\footnote{ASCA GOF calibration memo at  \\
http://lheawww.gsfc.nasa.gov/~yaqoob/ccd/nhparam.html}.

We then fitted each 
of the eight data sets (six \asca and two \rxte)
with a simple power-law model (including Galactic
absorption) and
a Gaussian to model the line,
utilizing all the data
in the selected energy bands ({\it ASCA}: 1.5--10 keV, 
PCA: 3--10 keV).
There were a maximum of three free
parameters associated with the line; 
the line center energy ($E_{c}$),
intrinsic Gaussian width ($\sigma_{\rm Fe}$), and intensity
($I_{\rm Fe}$). 
We found that all except
observations D/ PCA and E/ \asca had insufficient
signal-noise to constrain both the center energy
and intrinsic width simultaneously
so we fixed $\sigma_{\rm Fe}$  at 0.8 keV (consistent with the
best fitting values from observations D/ PCA and E/ \asca).
In cases where the line was particularly weak we also had
to fix $E_{c}$ (at 6.67 keV, from the best-constrained
measurement which was from observation E/\asca).
The spectral fitting results are shown in Table 2, which also shows the values
of $\Delta \chi^{2}$ compared to the case when no \fekalfa line 
was included in the model. Clearly, the line is detected with a 
very high level of confidence ( for the two {\it worst} cases
it is
$>98.2\%$  and $>99.975\%$ for observations B and F
respectively).
Note that in observations D/\asca and E/PCA the line is actually
resolved at the 90\% confidence level.
The best-fitting line energies from \asca are 
all $\sim 6.6$ keV, and the weighted mean, from the \asca observations
that could constrain the energy, is $6.62 \pm 0.20$ keV.
The
statistical uncertainty is still sufficiently large that
all ionization states of Fe except H-like are allowed.

Previous work has shown that a Compton-reflection continuum, common
in Seyfert 1 galaxies, is particularly weak in 3C 273 (e.g. Turner 
\etal 1990; Grandi \etal 1997). Nevertheless,
we tested whether Compton-reflection could affect the deduced width of
the \fekalfa line. Using the \asca data with the best-determined 
line width (observation E), we added a Compton-reflection continuum
to the power-law plus Gaussian model. For the most conservative 
constraints we assumed the primary X-ray continuum illuminates a
Compton-thick disk viewed face-on. The only additional free parameter
is the effective covering factor of the disk, $R$, as a fraction of
of the solid angle $2\pi$. The best-fitting value of $R$ was 0.0, with
a 90\% confidence upper limit of 0.20 (four interesting parameters,
or $\Delta \chi^{2} = 7.779$). Thus, the best-fitting line parameters
remained unchanged, with the bounds on the line width slightly larger,
($\sigma_{\rm Fe} = 0.80 \pm 0.51$ keV). 
Thus, Compton reflection does not affect our conclusion about the
broadness  of the \fekalfa line.
We used the PCA data
for observation E to obtain even tighter constraints on Compton-reflection.
Using the same model, and extending the PCA data up to 20 keV, we
obtained $R= 0.0$ with an upper limit of 0.094 (90\% confidence,
$\Delta \chi^{2} = 6.251$; recall that the line width had to be fixed
for the PCA/E data). 

As can be seen from Table 2, 
the smallest and largest measurements of $I_{\rm Fe}$ are
for observations A and F respectively and do not overlap
at the 90\% confidence level, yet these observations have 2--10 keV
fluxes which differ by only $\sim 10\%$.
Observations A and F also have the smallest and largest  
EW respectively,
which are also mutually exclusive at the 90\% confidence level.
However, a clear trend in the line variability is not apparent from these data;
studies with {\it XMM} will clarify this issue.

Since the broad \fekalfa emission line (Figures 1 and 2) is similar to that
seen in many Seyfert 1 galaxies (e.g. Nandra \etal 1997a; Reynolds
1997) we fitted the \asca data 
with
a model in which the line photons are
emitted in a disk rotating around a central Schwarzschild black hole
(e.g. Fabian \etal 1989). The parameters are
$\theta_{\rm obs}$ (inclination angle of the disk normal relative to
the observer),
$r_{i}$ (inner radius),
$r_{o}$ (outer radius), $q$ (power-law index characterizing the
line emissivity as $\propto r^{-q}$),
$I_{\rm D}$ (intensity), and $E_{D}$ (line energy in the
disk rest frame). The fits were insensitive to $r_{i}$ and
$r_{o}$ so they were 
fixed at $6  r_{g}$ and $1000  \ r_{g}$ respectively
($r_{g} \equiv GM/c^{2}$). The data could not constrain $q$ so
we fixed it at 2.5, a value consistent with the mean found from
a sample of Seyfert galaxies (e.g. Nandra \etal 1997a).
Unfortunately there
is too much interplay between $E_{D}$ and the other line parameters
so we fixed $E_{D}$ at 6.4 keV, corresponding to a cold disk.
Thus there are at most two free parameters in this model of the
\fekalfa line ($I_{D}$ and $\theta_{\rm obs}$). Moreover, we found that
only observations D and E could constrain $\theta_{\rm obs}$,
so for the other fits we fixed $\theta_{\rm obs}$ at the
best-fitting value for observation D. The
results are shown in Table 3. Both observations
D and E give a consistent value for $\theta_{\rm obs}$ of 
$60^{\circ}$, with a lower bound of $\sim 45^{\circ}$ (the upper
value is not bounded). This is somewhat surprising because
quasars such as 3C 273 are thought to be observed face-on (or near 
face-on). To investigate whether a higher rest-frame energy of the
\fekalfa line can lower the inclination angle, we repeated the 
spectral fits with the rest energy fixed at the
extreme value of 6.97 keV (corresponding 
to H-like Fe) and obtained $\theta_{\rm obs} = 46^{+15}_{-27}$ and
$22^{+28}_{-22}$ degrees for observations D and E respectively.
However, the fits are slightly {\it worse}, with $\Delta \chi^{2}$
between 2--3. 
Thus we cannot unambiguously constrain
the inclination angle and require better data, from {\it XMM} for example. 
Note that EW values from the disk model 
(for both values of the rest-frame energy) are virtually the same
as those obtained from the Gaussian model. It should also be pointed
out that in no case does the disk-line model formally provide a better
fit than the Gaussian model (but the differences in $\chi^{2}$ are
not significant).

\section{CONCLUSIONS}

From an analysis of the entire database of \asca observations of
3C 273 and two simultaneous \rxte observations, we have shown
that the \fekalfa line in this quasar
is very broad. 
The line is resolved in two of the observations with a mean
Gaussian width of $0.8 \pm 0.3$ keV, corresponding to a FWHM of
$\sim 0.3 \pm 0.1 c$. 
The 2--10 keV intrinsic luminosity ranged from
$\sim 0.8-2 \times \ 10^{46} \rm \ erg \ s^{-1}$. 
The smallest and largest equivalent widths we measured are 
$43^{+34}_{-34}$ eV and $133^{+52}_{-53}$ eV respectively (quasar frame)
but the data do not allow us to identify any clear trend of 
line variability and its relation to the continuum luminosity.
The broad \fekalfa line is similar to the
relativistic \fekalfa lines commonly found in Seyfert~1 galaxies,
which are
thought to be shaped by gravitational and Doppler shifts in the
vicinity of a black hole. The EW of the lines in Seyfert~1 galaxies is
higher however, typically $\sim 200-300$ eV.
The \fekalfa emission line
was previously thought 
to be narrow (FWHM $<10,000$ km/s) or absent 
in 
AGN with a 2--10 keV
luminosity exceeding $\sim 10^{45} \rm \ erg \ s^{-1}$.
Thus,
3C 273 is now the highest luminosity quasar in which a broad \fekalfa line
has been found. 
When modeled by a Gaussian,
the weighted mean peak energy of the line 
is $6.62 \pm 0.20$ keV, indicative of He-like Fe, although the
90\% confidence errors allow all lower ionization states.
If we model the \fekalfa line in 3C 273 as originating in an
accretion disk around a black hole, we obtain a disk inclination angle
$> 44^{\circ}$ if the \fekalfa line originates in cold matter.
If the line-emitting region is highly ionized the 
disk inclination angle is allowed to be smaller, including near face-on
for H-like Fe. 
The disk model also predicts that the 
\fekalfa line should be accompanied by a Compton-reflection continuum,
but we constrained the magnitude of the latter to be less than
10\% of that expected from a face-on, semi-infinite, Compton-thick disk
centrally illuminated by the primary X-ray continuum. 
The EW during some of the observations is too large compared to
that predicted for the observed reflection continuum (e.g. see George
\& Fabian 1991). Time lags longer than several
days between line and continuum and/or an
over-abundance of Fe could reduce this discrepancy. 
Time-resolved spectroscopy with {\it XMM} is needed to make further progress.

The authors thank the \asca mission operations team at ISAS, Japan,
and all the instrument teams 
for their dedication and hard work in making these \asca observations 
possible. The authors also thank Paola Grandi for several years of
fruitful collaborative work on 3C 273 and
{\it ASCA}/{\it BeppoSAX} cross-calibration,
T. Dotani for helpful comments,
Keith Jahoda for supplying up-to-date response matrices for the
PCA, and the \rxte GOF for help with data analysis issues. 
Finally, we thank an anonymous referee for making some important
suggestions.
This research made use of the HEASARC archives at the
Laboratory for High Energy Astrophysics, NASA/GSFC.

\newpage
\small
\begin{deluxetable}{lcccc}
\tablecaption{3C 273 Observation Log}
\tablecolumns{5}
\tablewidth{0pt}
\tablehead{
\colhead{Obs $^{a}$} & \colhead{START (UT)} & \colhead{END (UT)} & \colhead{Exposure
(ks)} & \colhead{Count Rate (ct/s)}
}
\startdata

A/SIS0 & 08/06/93 19:09 & 07/06/93 23:30 & 31.3 & 4.23 \nl

B/SIS0 & & & & \nl
(1) & 15/12/93 12:17 & 15/12/93 23:00 & 16.4 & 5.44 \nl
(2) & 15/12/93 23:53 & 16/12/93 12:10 & 19.0 & 5.12 \nl
(3) & 19/12/93 23:45 & 20/12/93 10:08 & 17.6 & 4.70 \nl
(4) & 20/12/93 10:45 & 20/12/93 21:29 & 18.2 & 4.89 \nl
TOTAL & 15/12/93 12:17 & 20/12/93 21:29  & 71.2 & 4.91 \nl

C/SIS0 & & & & \nl
(1) & 16/12/93 12:11 & 16/12/93 19:57 & 11.7 & 5.63 \nl
(2) & 20/12/93 22:04 & 21/12/93 04:00 & 10.2 & 4.76 \nl
(3) & 23/12/93 23:35 & 24/12/93 05:05 & 10.4 & 4.80 \nl
(4) & 27/12/93 10:37 & 27/12/93 17:59 & 10.6 & 5.71 \nl
TOTAL & 16/12/93 12:11 & 27/12/93 17:59  & 42.9 & 5.23 \nl

D/SIS0 &  17/7/96 20:05 & 18/7/96 13:32 & 41.9 & 2.44 \nl
D/PCA$^{b}$ &    17/7/96 23:17 & 18/7/96 13:52 & 47.3 & 22.1  \nl

E/SIS0 & 24/06/98  08:30 & 26/06/98 10:00 & 73.0 & 3.67 \nl
E/PCA$^{b}$ &    24/06/98 07:45 & 26/06/98 08:43 & 16.1 & 35.0  \nl

F/SIS0 & 09/01/00 16:35 & 10/01/00 23:32 & 45.2 & 3.69 \nl

\tablecomments{All count rates are background-subtracted. \\
$^{a}$ Although all four \asca instruments are
used in this analysis, we report here only the start/stop times,
exposure times and 0.5--10 keV count rates for the SIS0 spectra. \\
$^{b}$ The \rxte PCA count rates refer to the sum of PCUs 0,1 and 2,
layer 1, left and right combined, in the 2--10 keV band 
(see e.g. Weaver \etal 1998 for details). \\
} 
\enddata
\end{deluxetable}

\newpage
\small
\begin{deluxetable}{lcccccccc}
\tablecaption{Spectral Fits to 3C 273 with a Gaussian Line Model}
\tablecolumns{9}
\tablewidth{0pt}
\tablehead{
\colhead{Obs} & \colhead{$\Gamma$} & \colhead{$E_{c}$ (keV)} &
\colhead{$\sigma_{\rm Fe}$ (keV)} & \colhead{$I_{\rm Fe}$} &
\colhead{$EW$ (eV)} & \colhead{$F^{d}$} &  \colhead{$\chi^{2}$/ d.o.f.
$^{e}$} & \colhead{$\Delta \chi^{2}$ $^{f}$} 
}

\startdata

A/ \asca $^{b}$ & 
	$1.646^{+0.110}_{-0.110}$ & 	$6.62^{+0.36}_{-0.36}$ &
	$0.8$f  		 & 	$2.10^{+0.82}_{-0.84}$ &
	$133^{+52}_{-53}$ &	12.4	&	1643.4/1542	& 28.4  \nl

B/ \asca $^{a}$ &
        $1.570^{+0.007}_{-0.007}$ &      6.67f  &
        $0.8$f                   &      $1.18^{+0.53}_{-0.52}$ &
        $59^{+27}_{-25}$ &     16.8    &       1467.2/1891     & 13.7  \nl

C/ \asca $^{b}$ &
        $1.582^{+0.008}_{-0.007}$ &      $6.54^{+0.44}_{-0.41}$  &
        $0.8$f                   &      $1.61^{+0.75}_{-0.76}$ &
        $79^{+37}_{-37}$ &     16.3    &       1857.9/1757     & 21.2  \nl

D/ \asca $^{b}$ &
        $1.665^{+0.012}_{-0.012}$ &      $6.64^{+0.39}_{-0.38}$  &
        $0.8$f                   &      $1.24^{+0.51}_{-0.52}$ &
        $147^{+60}_{-61}$ &     6.9    &       1522.3/1485     & 26.3  \nl

D/ {\it PCA}  $^{c}$ &
        $1.646^{+0.010}_{-0.011}$ &      $6.35^{+0.24}_{-0.45}$  &
        $0.74^{+0.51}_{-0.33}$   &      $1.36^{+1.11}_{-0.45}$ &
        $116^{+95}_{-38}$ &     8.7    &       7.8/13     	& 150.6  \nl

E/ \asca $^{c}$ &
        $1.650^{+0.008}_{-0.008}$ &      $6.67^{+0.44}_{-0.41}$  &
        $0.80^{+0.42}_{-0.41}$   &      $1.09^{+0.79}_{-0.61}$ &
        $85^{+61}_{-47}$ &     10.7    &       1679.0/1678     & 27.5  \nl

E/ {\it PCA}  $^{b}$ &
        $1.675^{+0.011}_{-0.012}$ &      $6.43^{+0.46}_{-0.45}$  &
        $0.8$f		         &      $1.11^{+0.49}_{-0.48}$ &
        $61^{+27}_{-27}$ &     13.7    &       17.0/14           & 25.8  \nl

F/ {\it ASCA}  $^{a}$ & $1.675^{+0.008}_{-0.009}$ & 6.67f & 0.8f & 
	$0.60^{+0.47}_{-0.48}$ & $43^{+34}_{-34}$ & 11.3 & 
	1747.0/1641 & 5.60 \nl

\tablecomments{The continuum model used is a simple power law
with photon index, $\Gamma$, plus Galactic absorption with
$N_{H} = 1.79 \times 10^{20} \rm \ cm^{-2}$, plus a Gaussian
emission line.
The emission line parameters
are all referred to the quasar frame ($z=0.158$); these
are the intrinsic width, $\sigma_{\rm Fe}$, the center energy, $E_{c}$,
the intensity ($\rm 10^{-4} \rm \  photons \  cm^{-2} \ s^{-1} $), $I_{\rm Fe}$,
and the equivalent width (EW). Fixed parameters are denoted by `f' beside
the fixed value. \\
Errors on $\Gamma$ are all 90\% confidence for one interesting parameter. \\
$^{a}$ Errors on the line parameters 
are 90\% confidence for one interesting parameter
($\Delta \chi^{2} = 2.706$). \\
$^{b}$ Errors on the line parameters
are 90\% confidence for two interesting parameters
($\Delta \chi^{2} = 4.605$). \\
$^{c}$ Errors on the line parameters
are 90\% confidence for three interesting parameters
($\Delta \chi^{2} = 6.251$). \\
$^{d}$  Flux in the 2--10 keV {\it observed}
band,  in units of $10^{-11} \rm \ erg \ cm^{-2} \ s^{-1}$
(four-instrument average for {\it ASCA}). \\
$^{e}$ d.o.f. $=$ degrees of freedom. \\ 
$^{f}$ The decrease in $\chi^{2}$ when the Gaussian emission line is
added to the simple power-law plus Galactic absorption model only. 
}
\enddata
\end{deluxetable}

\newpage
\normalsize
\begin{deluxetable}{lcccc}
\tablecaption{Spectral Fits to 3C 273 ASCA Data with a 
Schwarzschild Disk Line Model}
\tablecolumns{5}
\tablewidth{0pt}
\tablehead{
\colhead{Obs} & \colhead{$\theta_{\rm obs}$ (degrees)} & 
\colhead{$I_{\rm D}$} &
\colhead{$EW$ (eV)} & \colhead{$\chi^{2}$/ d.o.f.
}  
}
\startdata

A & 59 f		& $1.73^{+0.56}_{-0.56}$ & 
	$130^{+42}_{-42}$ 	& 1645.4/1543 \nl
 
B & 59 f                & $0.76^{+0.46}_{-0.48}$ &
        $69^{+42}_{-44}$      & 1473.7/1891 \nl

C & 59 f                & $1.33^{+0.51}_{-0.50}$ &
        $76^{+30}_{-30}$      & 1859.7/1758 \nl

D & $59^{+31}_{-15}$                 & $1.05^{+0.88}_{-0.51}$ &
        $141^{+118}_{-68}$      & 1523.5/1485 \nl

E & $61^{+29}_{-12}$                 & $0.95^{+0.56}_{-0.38}$ &
        $82^{+49}_{-32}$      & 1681.2/1679 \nl

F & 59 f 		& $0.51^{+0.41}_{-0.42}$ &
	$42^{+34}_{-34}$ 	& 1747.3/1641 \nl

\tablecomments{Spectral fits to the \asca 3C 273 data with
the same continuum model as in Table 1, but replacing the Gaussian
model of the \fekalfa emission line with a relativistic line from
a disk around a Schwarzschild black hole (see text). The continuum
parameters obtained are virtually identical to those in Table 1.
The disk inclination is $\theta_{\rm obs}$. The line intensity
($I_{\rm D}$) and equivalent width ($EW$) are both referred to the
quasar frame ($z=0.158$). Errors are 90\% confidence for one
interesting parameter 
($\Delta \chi^{2} =2.706$) when $\theta_{\rm obs}$ is fixed (denoted by `f')
and for
two 
interesting parameters ($\Delta \chi^{2} =4.605$) otherwise. 
}
\enddata
\end{deluxetable}

\newpage
\section*{Figure Captions}

\par\noindent
{\bf Figure 1} \\
The \fekalfa line from the 1996 July campaign of 3C 273 (Obs D). The data were
fitted with a simple power-law plus Galactic absorption, omitting the
4--8 keV data (see text for details). Shown are the ratios of data 
compared to this model for \rxte PCA (top inset), \asca SIS0 (filled
circles) 
\asca SIS1 (crosses), \asca GIS2 and GIS3 (filled circles and
crosses, respectively, in bottom inset).  
The line is clearly broad, just like the profiles commonly 
found in Seyfert 1
galaxies.

\par\noindent
{\bf Figure 2} \\
All the data for 3C 273 taken by \asca over the entire mission
(1993--2000) and added together, separately for SIS0 and SIS1.
The total exposure time is $\sim 258$ ks.
Shown are the data/model ratios when the co-added spectra were
fitted with a power law, omitting the 4--8 keV band. The broad
\fekalfa line is a very prominent feature. Since the statistical errors
are so small, systematic problems with the SIS+XRT response show up
as a hard tail in SIS1 relative to SIS0.

\newpage

\begin{figure}[h]
\vspace{10pt}
\centerline{\psfig{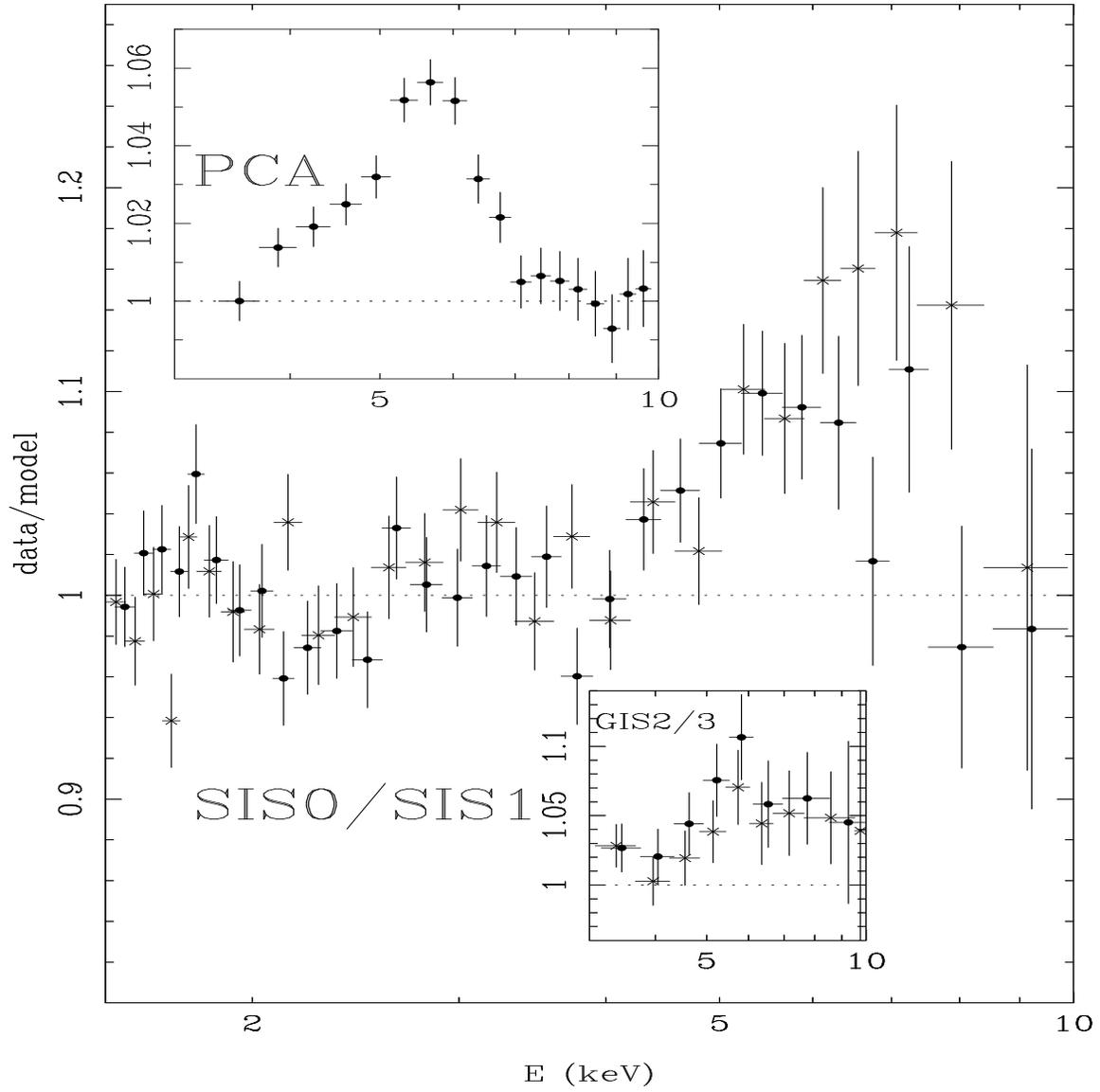}
}
\caption{See figure caption. }
\end{figure}

\begin{figure}[h]
\vspace{10pt}
\centerline{\psfig{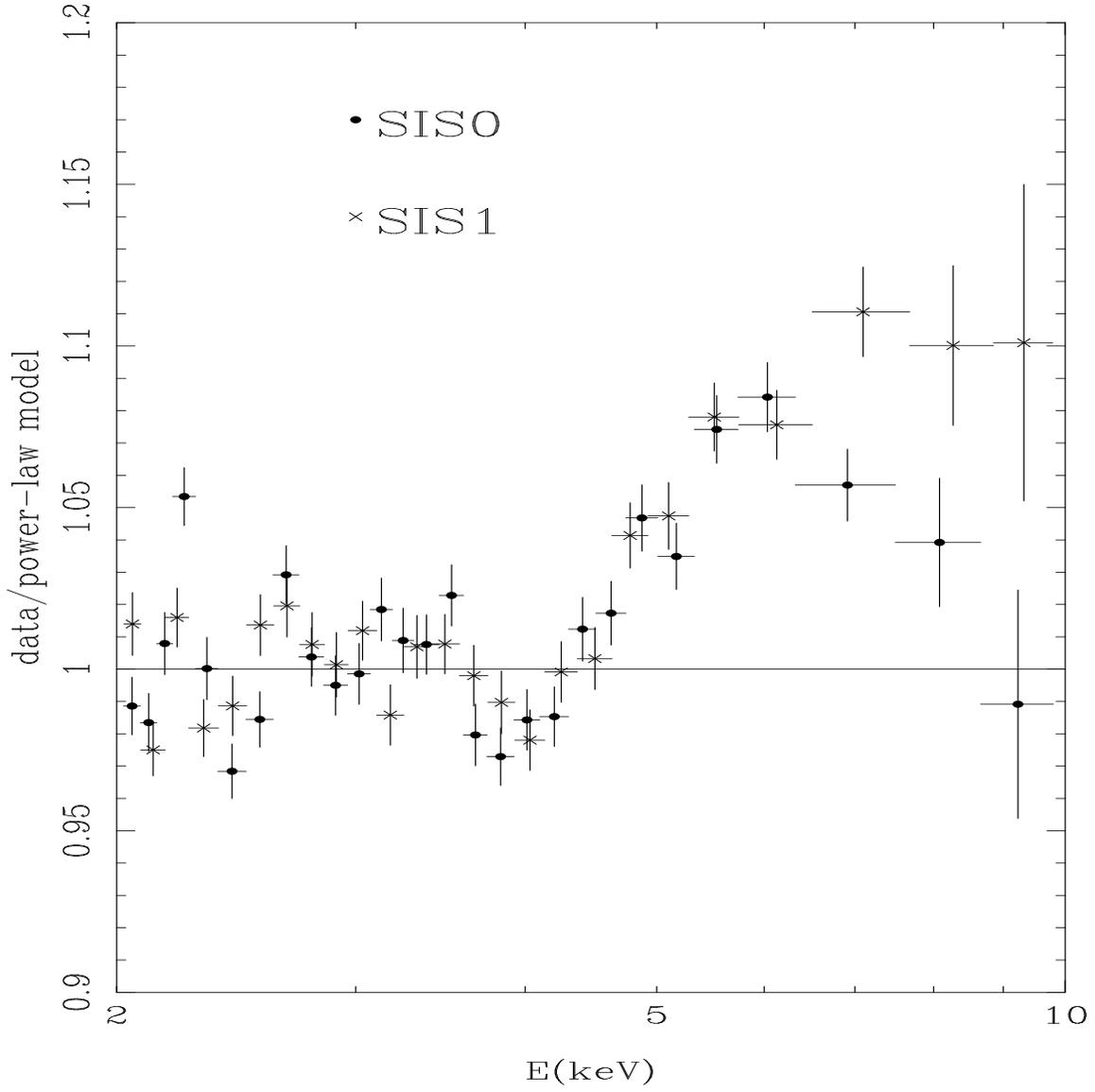}
}
\caption{See figure caption. }
\end{figure}

\end{document}